\documentclass[a4paper,12pt]{article}
\usepackage[english]{babel}
\usepackage{graphicx,color}
\usepackage{amsmath,amsthm,amssymb,amsfonts}
\usepackage{amsthm}
\usepackage{mathtext}
\usepackage{exscale}
\usepackage[T1]{fontenc}
\usepackage[cp1250]{inputenc}

\title{Bosonic Quantum Gravity According to the Global One-Dimensionality Conjecture}

\author{\textbf{Lukasz Andrzej Glinka}\footnote{E-mail to: laglinka@gmail.com, lukaszglinka@wp.eu}}
\date{\empty}

\begin{document}

\maketitle
\thispagestyle{empty}
\begin{abstract}
In this paper, making use of the global one-dimensionality conjecture, we discuss the reduction of the Wheeler--DeWitt quantum geometrodynamics to the Klein--Gordon equation describing the scalar bosonic particle. The method of second quantization in the appropriate Fock space is applied, and, employing both the Bogoliubov transformation as well as Heisenberg equations of motion, the quantum gravity is expressed as evolution of the creators and annihilators related to the initial data. It is shown that this procedure leads to the understanding of the boson mass, through the one-point two-boson quantum correlations, as a scaled initial data mass.\\

\end{abstract}
\newpage
\section{Wheeler-DeWitt Geometrodynamics}
Let us consider the Einstein--Hilbert field equations \cite{eh1,eh2} \footnote{Here the system of units $c=\hbar=k_B=8\pi G/3=1$ is applied.}
\begin{equation}\label{feq}
R_{\mu\nu}-\dfrac{R[g]}{2}g_{\mu\nu}+\Lambda g_{\mu\nu}=3T_{\mu\nu},
\end{equation}
where $g_{\mu\nu}$ is a non-degenerate and symmetric $\left(\!\!\!\begin{array}{c}0\vspace*{-4pt}\\2\end{array}\!\!\!\right)$-tensor field. Recall that the metric-contracted Riemann curvature tensor $R_{\mu\nu}$, the Einstein cosmological constant $\Lambda$, and the Ricci scalar curvature $R[g]=g^{\kappa\lambda}R_{\kappa\lambda}$ describe the differential geometry of a pseudo-Riemannian (Lorentzian) space-time manifold $(M,g)$ \cite{rie, kri}, whereas a stress-energy tensor $T_{\mu\nu}$ describes the Matter fields. The field equations of General Relativity (\ref{feq}) can be formulated as the result of the the least action principle
\begin{equation}\label{eh1}
\dfrac{\delta S[g]}{\delta g_{\mu\nu}}=0,
\end{equation}
which in this context known as the Palatini principle \cite{pal}, applied to the Einstein--Hilbert action supplemented by the boundary term
\begin{equation}\label{eh0}
S[g]=\int_{M}d^4x\sqrt{-g}\left\{-\dfrac{1}{6}R[g]+\dfrac{\Lambda}{3}+\mathcal{L}\right\}-\dfrac{1}{3}\int_{\partial M}d^3x\sqrt{h}K[h],
\end{equation}
where one allows variations for which the normal derivatives on $\partial M$ are non-zero, in order to cancel surface terms. Here $K[h]$ is the extrinsic curvature of an induced three-dimensional spacelike boundary $(\partial M,h)$ described by a metric $h_{ij}$, and $\mathcal{L}$ is the Lagrangian of Matter fields provoking the stress-energy tensor $T_{\mu\nu}$
 \begin{equation}
   T_{\mu\nu}=\frac{2}{\sqrt{-g}}\frac{\delta\left(\sqrt{-g}\mathcal{L}\right)}{\delta g^{\mu\nu}}.
 \end{equation}
Stationarity of the Matter fields results in a global timelike Killing vector field for a metric field $g_{\mu\nu}$. A coordinate system can be chosen such that the Killing vector field equals $\dfrac{\partial}{\partial t}$ and the foliation $t=constant$ is spacelike. Then, a metric field depends at most on a spatial coordinates $x^i$, the $t$ becomes a global time coordinate \cite{qft}, and the $3+1$ decomposition of a space-time metric
\begin{eqnarray}\label{dec}
  &&g_{\mu\nu}=\left[\begin{array}{cc}-N^2+N_iN^i&N_j\\N_i&h_{ij}\end{array}\right],~~g^{\mu\nu}=\left[\begin{array}{cc}-\dfrac{1}{N^2}&\dfrac{N^j}{N^2}\vspace*{5pt}\\ \dfrac{N^i}{N^2}&h^{ij}-\dfrac{N^iN^j}{N^2}\end{array}\right],\\
  &&h_{ik}h^{kj}=\delta_i^j,~~N^i=h^{ij}N_j,~~g=N^2h,
\end{eqnarray}
has also a global sense. In this case, the action (\ref{eh0}) becomes
\begin{eqnarray}\label{gd}
  S[g]\!\!\!\!&=&\!\!\!\!\int dt~L(\pi,\pi^i,\pi^{ij},N,N_i,h_{ij}),\\
  L(\pi,\pi^i,\pi^{ij},N,N_i,h_{ij})\!\!\!\!&=&\!\!\!\!\int_{\partial M} d^3x\left\{\pi\dot{N}+\pi^i\dot{N_i}+\pi^{ij}\dot{h}_{ij}-NH-N_iH^i\right\},
\end{eqnarray}
where
\begin{eqnarray}
\dot{h}_{ij}&=&\dfrac{\partial h_{ij}}{\partial t}=N_{i|j}+N_{j|i}-2NK_{ij},\label{con0}\\
H&=&\sqrt{h}\left\{K^2-K_{ij}K^{ij}+R[h]-2\Lambda-6T_{nn}\right\},\label{con1}\\
H^i&=&-2\pi^{ij}_{~;j}=-2\pi^{ij}_{~,j}-h^{il}\left(2h_{jl,k}-h_{jk,l}\right)\pi^{jk},\label{con2}
\end{eqnarray}
where the second formula follows from the Gauss-Codazzi equations \cite{han}. Here $K_{ij}$ is the extrinsic-curvature tensor ($K=K^{i}_{i}$), and $\pi^{ij}$ is the canonical conjugate momentum field to the field $h_{ij}$
\begin{equation}\label{mom}
\pi^{ij}=\dfrac{\delta L}{\delta\dot{h}_{ij}}=-\sqrt{h}\left(K^{ij}-h^{ij}K\right).
\end{equation}
Time-preservation requirement \cite{dir1} of the primary constraints \cite{dew} for (\ref{gd})
\begin{equation}
  \pi=\dfrac{\delta L}{\delta\dot{N}}\approx0,~~\pi^i=\dfrac{\delta L}{\delta\dot{N_i}}\approx0,
\end{equation}
leads to the secondary constraints
\begin{equation}
  H\approx0,~~H^i\approx0,\label{const}
\end{equation}
called the Hamiltonian constraint and the diffeomorphism constraint, respectively. The diffeomorphism constraint merely reflects spatial diffeoinvariance, and the Hamiltonian constraint gives the dynamics. By (\ref{mom}) the Hamiltonian constraint becomes the Einstein--Hamilton--Jacobi equation (For more detailed derivations and discussions Cf. the Refs. \cite{ham1}-\cite{ham33})
\begin{equation}\label{con}
G_{ijkl}\pi^{ij}\pi^{kl}+\sqrt{h}\left(R[h]-2\Lambda-6T_{nn}\right)=0,
\end{equation}
where $G_{ijkl}$ is called the Wheeler superspace metric
\begin{equation}\label{ss}
G_{ijkl}=\dfrac{1}{2\sqrt{h}}\left(h_{ik}h_{jl}+h_{il}h_{jk}-h_{ij}h_{kl}\right).
\end{equation}
The primary quantization \cite{dir} of the constraints (\ref{con}) can be done through application of the canonical commutation relations \cite{fad}
\begin{eqnarray}\label{dpq}
i\left[\pi^{ij}(x),h_{kl}(y)\right]&=&\dfrac{1}{2}\left(\delta_{k}^{i}\delta_{l}^{j}+\delta_{l}^{i}\delta_{k}^{j}\right)\delta^{(3)}(x,y),\\
i\left[\pi^i(x),N_j(y)\right]&=&\delta^i_j\delta^{(3)}(x,y),\\
i\left[\pi(x),N(y)\right]&=&\delta^{(3)}(x,y),
\end{eqnarray}
and leads to the Wheeler--DeWitt equation \cite{whe, dew}
\begin{equation}\label{wdw}
\left\{-G_{ijkl}\dfrac{\delta^2}{\delta h_{ij}\delta h_{kl}}+h^{1/2}\left(R[h]-2\Lambda-6T_{nn}\right)\right\}\Psi[h_{ij}]=0,
\end{equation}
while the other first class constraints are merely auxiliarly conditions on the wave function $\Psi[h_{ij}]$
\begin{equation}
  \pi\Psi[h_{ij}]=0,~~\pi^i\Psi[h_{ij}]=0,~~H^i\Psi[h_{ij}]=0.
\end{equation}
Furthermore, the following canonical commutation relations hold
\begin{equation}
\left[\pi(x),\pi^i(y)\right]=\left[\pi(x),H^i(y)\right]=\left[\pi^i(x),H^j(y)\right]=\left[\pi^i(x),H(y)\right]=0,
\end{equation}
and, in consequence, $H_i$ are generators of diffeomorphisms $\widetilde{x}^i=x^i+\delta x^i$ \cite{dew}
\begin{eqnarray}
\left[h_{ij},i\int_{\partial M}H_{a}\delta x^a d^3x\right]&=&-h_{ij,k}\delta x^k-h_{kj}\delta x^{k}_{~,i}-h_{ik}\delta x^{k}_{~,j}~~,\\
\left[\pi_{ij},i\int_{\partial M}H_{a}\delta x^a d^3x\right]&=&-\left(\pi_{ij}\delta x^k\right)_{,k}+\pi_{kj}\delta x^{i}_{~,k}+\pi_{ik}\delta x^{j}_{~,k}~~,
\end{eqnarray}
or in more conventional form
\begin{eqnarray}
  i\left[H_i(x),H_j(y)\right]\!\!&=&\!\!\int_{\partial M}H_{a}c^a_{ij}d^3z,\label{com1}\\
  i\left[H(x),H_i(y)\right]\!\!&=&\!\!H\delta^{(3)}_{,i}(x,y),\label{com2}\\
  i\left[\int_{\partial M}H\delta x_1d^3x,\int_{\partial M}H\delta x_2d^3x\right]\!\!&=&\!\!\int_{\partial M}H^a\left(\delta x_{1,a}\delta x_2-\delta x_1\delta x_{2,a}\right)d^3x,\label{com3}
\end{eqnarray}
where $H_i=h_{ij}H^j$, and
\begin{equation}
  c^a_{ij}=\delta^a_i\delta^b_j\delta^{(3)}_{,b}(x,z)\delta^{(3)}(y,z)-\delta^a_j\delta^b_i\delta^{(3)}_{,b}(y,z)\delta^{(3)}(x,z)~~,
\end{equation}
are structure constants of diffeomorphism group. Commutators  (\ref{com1}-\ref{com3}) show the first-class constrained system property.

 Since the late 1960s, that is more than thirty years, quantum geometrodynamics based on the Wheeler--DeWitt equation (\ref{wdw}) has been studied with a quite well intensity, Cf. the Refs. \cite{qgr1}-\cite{qgr29}, but the general way to its solution is still unknown, and the only simplest situations merely related to cosmology are well described. This equation is an evolutionary equation on $S(\partial M)$, well known as the Wheeler superspace \cite{sup}, which is defined as a space of all equivalence class of metrics related by the action of the diffeomorphism group of a compact, connected, orientable, Hausdorff, $C^\infty$ 3-dimensional spacelike manifold without a boundary $\partial M$. Defining $Riem(\partial M)$ as consisting of all $C^\infty$ Riemannian metrics on $\partial M$, and $Dif\!f(\partial M)$ as a group of all $C^\infty$ diffeomorphisms of $\partial M$ that preserve orientation, the Wheeler superspace $S(\partial M)$ becomes the space of all orbits of $Dif\!f(\partial M)$, that is formally $S(\partial M)=Riem(\partial M)/Dif\!f(\partial M)$. $S(\partial M)$ is a connected, second-countable, metrizeable space. All geometries with the same kind of symmetry are manifold in $S(\partial M)$, that is they have homeomorphic neighbouhoods. However, symmetric geometries' neighbourhoods are not homeomorphic to nonsymmetric geometries' ones, and, for this reason, $S(\partial M)$ is not a manifold. Superspace can be decomposed by its subspaces on a countable, partially-ordered, $C^\infty$-Fr$\mathrm{\acute{e}}$chet manifold partition, that is an inverted stratification indexed by the symmetry type, or, in other words, geometries with a given symmetry are completely contained within the boundary of less symmetric geometries. The mostly studied situations are the so-called minisuperspace models which define purely quantum cosmological part of the Wheeler-DeWitt equation, Cf. the Refs. \cite{min1}-\cite{min7}, which formally study a certain specific strata of the superspace. Fischer \cite{sup} proved that through both a suitable choice of a subgroup of $Dif\!f(\partial M)$ and action of this subgroup on $Riem(\partial M)$, for $n$-dimensional boundary $\partial M$ the superspace $S(\partial M)$ can be extended to a manifold $S_e(\partial M)$ such that $\dim S_e(\partial M)/S(\partial M)=n(n+1)$.

\section{Global Bosonization}
The formulation of quantum gravity throughout the concept of the Wheeler superspace has no known physical consequences \cite{ish}, but is the crucial structural problem of quantum geometrodynamics. In this section we will present the construction of an appropriate linearization of Wheeler-Dewitt geometrodynamics, which we call here the bosonization in the global one-dimension.
\subsection{Reduction Problem}
Let us consider the standard relation of General Relativity \cite{mtw} between the variations of a metric field and its determinant
\begin{equation}\label{dg}
  \delta g = gg^{\mu\nu}\delta g_{\mu\nu}=g\left(g^{00}\delta g_{00}+g^{ij}\delta g_{ij}+g^{0j}\delta g_{0j}+g^{i0}\delta g_{i0}\right).
\end{equation}
The $3+1$ parametrization (\ref{dec}) allows determine the partial variations
\begin{eqnarray}
  \delta g_{00}&=&-\delta N^2+N^iN^j\delta h_{ij}+h_{ij}N^i\delta N^j+h_{ij}N^j\delta N^i,\\
  \delta g_{ij}&=&\delta h_{ij},\\
  \delta g_{0j}&=&h_{ij}\delta N^i+N^i\delta h_{ij},\\
  \delta g_{i0}&=&h_{ij}\delta N^j+N^j\delta h_{ij},
\end{eqnarray}
as well as the total variation
\begin{eqnarray}
\delta g=N^2\delta h+h\delta N^2.
\end{eqnarray}
Taking a contravariant metric field components of (\ref{dec}), and making use of (\ref{dg}), one obtains
\begin{eqnarray}
  N^2\delta h=N^2hh^{ij}\delta h_{ij},
\end{eqnarray}
and, therefore, one can easily establish the global relation between the first functional derivatives
\begin{equation}\label{red}
  \dfrac{\delta}{\delta h_{ij}} = hh^{ij}\dfrac{\delta}{\delta h},
\end{equation}
which, in fact, is the global reduction (\ref{red}) having a deep geometrical sense. Namely, from this point of view, the first functional derivative operator $\dfrac{\delta}{\delta h_{ij}}$ is an object from a vector space spanned by the contravariant three-dimensional metric $h^{ij}$ and, in consequence of the formula (\ref{red}), one can determine the Wheeler--DeWitt second derivative functional operator (\ref{wdw}) as follows
\begin{eqnarray}
-G_{ijkl}\dfrac{\delta^2}{\delta h_{ij}\delta h_{kl}}=\dfrac{3}{2}h^{3/2}\dfrac{\delta^2}{\delta h^2},
\end{eqnarray}
where we have used the identity
\begin{eqnarray}
  \left(h_{ik}h_{jl}+h_{il}h_{jk}-h_{ij}h_{kl}\right)h^{ij}h^{kl}=\delta_i^l\delta^i_l+\delta^j_l\delta^l_j-\delta^i_i\delta^k_k=-3.
\end{eqnarray}
For full consistency, one should also consider $\Psi[h]$ rather than $\Psi[h_{ij}]$ as the appropriate geometrodynamical wave function, what we call the global one-dimensionality conjecture. Hence the Wheeler--DeWitt equation (\ref{wdw}) transforms into the one-dimensional Klein--Gordon equation
\begin{equation}\label{new}
\left(\dfrac{\delta^2}{\delta{h^2}}+m^2\right)\Psi[h]=0,
\end{equation}
where
\begin{equation}\label{masqr}
m^2\equiv m^2[h]=\dfrac{2}{3h}\left(R[h]-2\Lambda-6T_{nn}\right),
\end{equation}
is the square of mass of the bosonic field $\Psi[h]$. Making use of the notation
\begin{equation}
  \Phi=\left[\begin{array}{c}\Psi\\ \Pi_\Psi\end{array}\right],~~\vec{\partial}=\left[\begin{array}{c}\dfrac{\delta}{\delta h}\\0\end{array}\right],~~\mathbb{M}=\left[\begin{array}{cc}
0&1\\-m^{2}&0\end{array}\right]\geq0,
\end{equation}
the second order scalar equation (\ref{new}) becomes the first order vector equation
\begin{equation}\label{dir}
 \left(i\mathbf{\Gamma}\vec{\partial}-\mathbb{M}\right)\Phi[h]=0,
\end{equation}
where $\Gamma$ matrices obey the relations
\begin{equation}
  \mathbf{\Gamma}=\left[-i\mathbf{1},\mathbf{0}\right],~~\left\{\mathbf{\Gamma}^{a},\mathbf{\Gamma}^{b}\right\}=2\eta^{ab}\mathbf{1},~~\eta^{ab}=\left[\begin{array}{cc}-1&0\\0&0\end{array}\right],
\end{equation}
where $\mathbf{1}$ and $\mathbf{0}$ are unit and null two-dimensional matrices.

We have seen that application of the global reduction (\ref{red}) to the Wheeler--DeWitt equation (\ref{wdw}), that throughout a global character of the decomposition (\ref{dec}) has also a global nature, results in the quantum theory of the scalar boson (\ref{new}). This scalar-type second order functional evolution was reduced directly to the vector-type first order functional equation (\ref{dir}) with a certain two-component field $\Phi[h]$ as the solution. In the evolutionary equation (\ref{new}), as well as in its reduced form (\ref{dir}), the superspace metric is completely absent, what means that the reduced geometrodynamics avoids the concept of the Wheeler superspace, whereas all other properties of quantum gravity remain unchanged. For this reason, the most mysterious element of the logical way presented in the Wheeler-DeWitt quantum geometrodynamics has been formally excluded from considerations as at least unnecessary. Strictly speaking, the notion of the Wheeler superspace as well as its mathematical properties are not need for the further analysis of the quantum gravity. In a certain language, one can say that we have transformed a whole theory into its subtheory, what in the language of superspace can be called a strata. However, the reduction process also showed that the strata is equivalent to all theory, what makes this derivation a new finding. In further developments of this paper, we will concentrate on the canonical quantization in the bosonic Fock space of the reduced quantum geometrodynamics (\ref{dir}).

\subsection{Initial Data Operator Basis}
Next step of the procedure is the field quantization of the equation (\ref{dir})
\begin{equation}\label{qkg}
   \Phi[h]\rightarrow\mathbf{\Phi}[h] \Rightarrow \left(i\mathbf{\Gamma}\vec{\partial}-\mathbb{M}\right)\mathbf{\Phi}[h]=0,
\end{equation}
which, by the openly bosonic character of the theory, should be done according to canonical commutation relations proper for the Bose--Einstein statistics, Cf. the Refs. \cite{neu,a-w,bos}, that is
\begin{eqnarray}
  \left[\mathbf{\Pi}_{\Psi}[h'],\mathbf{\Psi}[h]\right]&=&-i\delta(h'-h),\label{c1}\\
  \left[\mathbf{\Pi}_{\Psi}[h'],\mathbf{\Pi}_{\Psi}[h]\right]&=&0,\label{c2}\\
  \left[\mathbf{\Psi}[h'],\mathbf{\Psi}[h]\right]&=&0.\label{c3}
\end{eqnarray}
Making use of the method of second quantization, Cf. the Refs. \cite{ber, bs, bog}, from the equation (\ref{new}) spring that the field operator $\mathbf{\Phi}[h]$ can be represented in the Fock space of annihilation and creation functional operators
\begin{equation}\label{sqx}
  \mathbf{\Phi}[h]=\mathbb{Q}[h]\mathfrak{B}[h],
\end{equation}
where $\mathfrak{B}[h]$ is a dynamical basis in the Fock space
\begin{equation}\label{db}
  \mathfrak{B}[h]=\left\{\left[\begin{array}{c}\mathsf{G}[h]\\
\mathsf{G}^{\dagger}[h]\end{array}\right]:\left[\mathsf{G}[h'],\mathsf{G}^{\dagger}[h]\right]=\delta\left(h'-h\right), \left[\mathsf{G}[h'],\mathsf{G}[h]\right]=0\right\},
\end{equation}
and $\mathbb{Q}[h]$ is the second quantization matrix
\begin{equation}\label{sqxm}
  \mathbb{Q}[h]=\left[\begin{array}{cc}\dfrac{1}{\sqrt{2|m[h]|}}&\dfrac{1}{\sqrt{2|m[h]|}}\\
-i\sqrt{\dfrac{|m[h]|}{2}}&i\sqrt{\dfrac{|m[h]|}{2}}\end{array}\right].
\end{equation}
In this way, the operator equation (\ref{qkg}) becomes the equation for the operator basis $\mathfrak{B}[h]$
\begin{equation}\label{df}
\dfrac{\delta\mathfrak{B}[h]}{\delta h}=\left[\begin{array}{cc}
-im[h]&\dfrac{1}{2m[h]}\dfrac{\delta m[h]}{\delta h}\\
\dfrac{1}{2m[h]}\dfrac{\delta m[h]}{\delta h}&im[h]\end{array}\right]\mathfrak{B}[h].
\end{equation}
Actually, there is a nonlinearity given by coupling between annihilation and creation operators present as nondiagonal terms in (\ref{df}) and, consequently, the equation (\ref{df}) can not be solved standardly. In order to find the solutions, let us suppose that in the Fock space exists a new basis $\mathfrak{B}^\prime[h]$
\begin{equation}\label{pr1}
  \mathfrak{B}^\prime[h]=\left\{\left[\begin{array}{c}\mathsf{G}^\prime[h]\\
\mathsf{G}^{\prime\dagger}[h]\end{array}\right]:\left[\mathsf{G}^\prime[h'],\mathsf{G}^{\prime\dagger}[h]\right]=\delta\left(h'-h\right), \left[\mathsf{G}^\prime[h'],\mathsf{G}^\prime[h]\right]=0\right\},
\end{equation}
wherein both the Bogoliubov transformation
\begin{equation}\label{pr2}
\mathfrak{B}^\prime[h]=\left[\begin{array}{cc}u[h]&v[h]\\
v^{\ast}[h]&u^{\ast}[h]\end{array}\right]\mathfrak{B}[h],~~|u[h]|^2-|v[h]|^2=1,
\end{equation}
as well as the Heisenberg canonical equations of motion
\begin{equation}\label{pr3}
\dfrac{\delta\mathfrak{B}^\prime[h]}{\delta h}=\left[\begin{array}{cc}
-i\lambda[h] & 0 \\ 0 &
i\lambda[h]\end{array}\right]\mathfrak{B}^\prime[h],
\end{equation}
hold together. The diagonalization procedure (\ref{pr1})-(\ref{pr3}) converts a whole evolution (\ref{df}) from the operator basis onto the Bogoliubov coefficients
\begin{equation}\label{bcof}
  \dfrac{\delta}{\delta h}\left[\begin{array}{c}u[h]\\v[h]\end{array}\right]=\left[\begin{array}{cc}
-im[h]&\dfrac{1}{2m[h]}\dfrac{\delta m[h]}{\delta h}\\
\dfrac{1}{2m[h]}\dfrac{\delta m[h]}{\delta h}&im[h]\end{array}\right]\left[\begin{array}{c}u[h]\\v[h]\end{array}\right],
\end{equation}
and $\mathfrak{B}^\prime[h]$ becomes a static operator basis associated with the initial data
\begin{equation}\label{in}
\mathfrak{B}^\prime[h]\equiv\mathfrak{B}_{I}=\left\{\left[\begin{array}{c}\mathsf{G}_I\\
\mathsf{G}^{\dagger}_I\end{array}\right]: \left[\mathsf{G}_I,\mathsf{G}^{\dagger}_I\right]=1, \left[\mathsf{G}_I,\mathsf{G}_I\right]=0\right\},
 \end{equation}
and the vacuum state is also static
\begin{equation}
|0\rangle_I=\left\{|0\rangle_I:\mathsf{G}_I|0\rangle_I=0,~0={_I}\langle0| \mathsf{G}_I^\dagger\right\}.
\end{equation}

In the other words, now the whole integrability problem of the theory is contained in the evolutionary equations (\ref{bcof}). Nevertheless, the Bogoliubov coefficients are additionally constrained by the hyperbolic rotation condition (\ref{pr1}) and, for this reason, this is useful to apply the so-called superfluid parametrization, for which the solutions are given ad hoc as follows
\begin{eqnarray}
u[h]&=&\dfrac{1+\mu[h]}{2\sqrt{\mu[h]}}\exp\left\{im_I\int_{h_I}^{h}\dfrac{\delta h'}{\mu[h']}\right\},\label{sup1}\\
v[h]&=&\dfrac{1-\mu[h]}{2\sqrt{\mu[h]}}\exp\left\{-im_I\int_{h_I}^{h}\dfrac{\delta h'}{\mu[h']}\right\},\label{sup2}
\end{eqnarray}
where $\mu[h]$ is a mass scale
\begin{equation}\label{sca}
  \mu[h]=\dfrac{m_I}{m[h]}.
\end{equation}
This establishes the relation between a dynamical basis $\mathfrak{B}[h]$ and the initial data basis $\mathfrak{B}_I$
\begin{equation}
  \mathfrak{B}[h]=\mathbb{G}[h]\mathfrak{B}_I,
\end{equation}
where the transformation matrix $\mathbb{G}[h]$ is
\begin{eqnarray}\label{mon}
\mathbb{G}[h]=\left[\begin{array}{cc}
\dfrac{\mu[h]+1}{2\sqrt{\mu[h]}}e^{-i\theta[h]}\vspace*{10pt}&
\dfrac{\mu[h]-1}{2\sqrt{\mu[h]}}e^{i\theta[h]}\\
\dfrac{\mu[h]-1}{2\sqrt{\mu[h]}}e^{-i\theta[h]}&
\dfrac{\mu[h]+1}{2\sqrt{\mu[h]}}e^{i\theta[h]}\end{array}\right],
\end{eqnarray}
where $i\theta[h]$ is given by a phase of (\ref{sup1}). By this reason, the solution of the equation (\ref{qkg}) can be expressed in the initial data basis as
\begin{equation}\label{phi}
  \mathbf{\Phi}[h]=\mathbb{Q}[h]\mathbb{G}[h]\mathfrak{B}_I.
\end{equation}
\subsection{One-point correlations}
The second quantized equation (\ref{new}) can be rewritten in the following form
\begin{equation}
\left(\mu^2[h]\dfrac{\delta^2}{\delta h^2}+m^2_I\right)\mathbf{\Psi}[h]=0,
\end{equation}
and its solution can be easily derived as a direct conclusion of (\ref{phi})
\begin{eqnarray}\label{field}
  \mathbf{\Psi}[h]=\frac{\mu[h]}{2\sqrt{2m_I}}\left(\exp\left\{-im_I\int_{h_I}^h\dfrac{\delta h'}{\mu[h']}\right\}\mathsf{G}_I+\exp\left\{im_I\int_{h_I}^h\dfrac{\delta h'}{\mu[h']}\right\}\mathsf{G}_I^\dagger\right).
\end{eqnarray}
This field acts on the initial data vacuum state as follows
\begin{eqnarray}
  \mathbf{\Psi}[h]|0\rangle_I&=&\frac{\mu[h]}{2\sqrt{2m_I}}e^{i\theta[h]}\mathsf{G}^{\dagger}_I|0\rangle_I,\\
  {_I}\langle0|\mathbf{\Psi}^\dagger[h]&=&{_I}\langle0|\mathsf{G}_I\frac{\mu[h]}{2\sqrt{2m_I}}e^{-i\theta[h]}.
\end{eqnarray}
and, for this reason, one can consider the following many-field quantum states
\begin{eqnarray}
|h,n\rangle&\equiv&\left(\mathbf{\Psi}[h]\right)^n|0\rangle_I=\left(\frac{\mu[h]}{2\sqrt{2m_I}}e^{i\theta[h]}\right)^n\mathsf{G}^{\dagger n}_I|0\rangle_I,\\
\langle n',h'|&\equiv&{_I}\langle0|\left(\mathbf{\Psi}^\dagger[h']\right)^{n'}={_I}\langle0|\mathsf{G}_I^{n'}\left(\frac{\mu[h']}{2\sqrt{2m_I}}e^{-i\theta[h']}\right)^{n'},
\end{eqnarray}
and determine the two-point quantum correlator of two many-field states
\begin{eqnarray}\label{gencor}
  \langle n',h'|h,n\rangle=\dfrac{\mu^{n'}[h']\mu^n[h]}{\left(8m_I\right)^{(n'+n)/2}}e^{-im_I\theta_{n',n}[h',h]}\langle0|\mathsf{G}_I^{n'}\mathsf{G}^{\dagger n}_I|0\rangle_I,\label{cor0}
\end{eqnarray}
where
\begin{equation}
\theta_{n',n}[h',h]=n'\int_{h_I}^{h'}\dfrac{\delta h''}{\mu[h'']}-n\int_{h_I}^{h}\dfrac{\delta h''}{\mu[h'']}.
\end{equation}
Application of the normalization
\begin{equation}\label{norm}
  \langle 1,h_I|h_I,1\rangle=\dfrac{1}{8m_I}{_I}\langle0|0\rangle_I\equiv1\Longrightarrow{_I}\langle0|0\rangle_I=8m_I,
\end{equation}
allows to define the following correlators
\begin{eqnarray}
  \langle n',h|h,n\rangle&=&\left(\dfrac{\langle 1,h|h,1\rangle}{{_I}\langle0|0\rangle_I}\right)^{(n'+n)/2}e^{-i(n'-n)\theta[h]}{_I}\langle0|\mathsf{G}_I^{n'}\mathsf{G}^{\dagger n}_I|0\rangle_I,\label{cor4}\\
  \dfrac{\langle n,h'|h,n\rangle}{{_I}\langle 0|0\rangle_I}&=&\left(\dfrac{\langle 1,h'|h,1\rangle}{{_I}\langle0|0\rangle_I}\right)^n,\label{cor3}
\end{eqnarray}
where
\begin{eqnarray}
\langle 1,h'|h,1\rangle&=&\mu[h']\mu[h]\exp\left\{im_I\int_{h'}^{h}\dfrac{\delta h''}{\mu[h'']}\right\},\label{cor1}\\
    \langle 1,h|h,1\rangle&=&\mu^2[h].\label{cor2}
\end{eqnarray}
The last formula (\ref{cor2}) together with the definition (\ref{sca}) leads to the relation between the mass of the bosonic field $\mathbf{\Psi}[h]$ and the initial data mass $m_I$
\begin{equation}\label{lam}
  m[h]=\lambda[h]m_I,~~\lambda[h]=\dfrac{1}{\sqrt{\langle 1,h|h,1\rangle}},
\end{equation}
that means the arbitrary mass $m[h]$ is only the the initial data mass $m_I$ scaled by a coefficient $\lambda[h]$, where $\lambda^2$ is a reciprocate of the one-point correlations of the quantum bosonic field $\mathbf{\Psi}[h]$. Therefore, actually for an arbitrary point $h$ the mass $m[h]$ is given through the correlations of two bosonic fields $\mathbf{\Psi}$ in this point, what establishes the crucial physical role of the superspace for quantum gravity.
\section{Summary}
Although making use of the Hamiltonian primary quantization of General Relativity, the procedure of global bosonization of the Wheeler--DeWitt quantum geometrodynamics and its second quantization essentially differs from the previous authors' results. The new aspect is quantum field theory formulation which leads to the initial data basis and considers quantum gravity as the theory of a scalar boson. The proposed approach is far from still implausible approach offered by high energy physics \cite{veltman} which makes gravitons the only spin 2 particles living in nature, the so-called third quantization, Cf. the Refs. \cite{th1}-\cite{th7}, where operator bases are not applied at all, and other physically unverifiable mathematical theories. The main goal of the presented approach is the physical role for the reduced geometrodynamics and its superspace through the one-point correlations and, moreover, formulation of quantum gravity through a scalar boson whose mass is the only scaled initial datum. A reader interested in the developments of this approach is advised to take into account the author's monograph \cite{glinka}.


\begin{thebibliography}{99}
\bibitem{eh1}   A. Einstein, \textrm{Sitzungsber. Preuss. Akad. Wiss. Berlin} \textbf{44}, N2, 778, (1915); \textit{ibid.} \textbf{46}, N2, 799, (1915); \textit{ibid.} \textbf{48}, N2, 844 (1915).
\bibitem{eh2}   D. Hilbert, \textrm{Konigl. Gesell. d. Wiss. G\"ottinger, Nachr., Math.-Phys. Kl.} \textbf{27}, 395 (1915).
\bibitem{rie}   B. Riemann, \textrm{Nachr. Ges. Wiss. G\"ottingen} \textbf{13}, 133 (1920).
\bibitem{kri}   M. Kriele, \textit{Spacetime. Foundations of General Relativity and Differential Geometry}, \textrm{Lect. Notes Phys. Monogr.} \textbf{59}, Springer-Verlag, Berlin Heidelberg New York, (1999).
\bibitem{pal}   A. Palatini, \textrm{Rend. Pal.} \textbf{43}, 203 (1919).
\bibitem{qft}   B. DeWitt, \textit{The Global Approach to Quantum Field Theory, Vol. 1,2}, \textrm{Int. Ser. Monogr. Phys.} \textbf{114}, Clarendon Press, Oxford (2003).
\bibitem{han}   A. Hanson, T. Regge, and C. Teitelboim, \textit{Constrained Hamiltonian Systems}, Contributi del Centro Linceo Interdisciplinare di Scienze Matematiche e loro Applicazioni, n. 22, Accademia Nazionale dei Lincei, Roma (1976).
\bibitem{dir1}  P.A.M. Dirac, \textrm{Lectures on Quantum Mechanics}, Belfer Graduate School of Science, Yeshiva University, New York (1964).
\bibitem{dew}   B.S. DeWitt, \textrm{Phys. Rev.} \textbf{160}, 1113 (1967).
\bibitem{ham1}  F.A.E. Pirani and A. Schild, \textrm{Phys. Rev.} \textbf{79}, 986 (1950).
\bibitem{ham2}  P.G. Bergmann, \textrm{Helv. Phys. Acta. Suppl.} \textbf{4}, 79 (1956); \textrm{Nuovo Cim.} \textbf{3}, 1177 (1956).
\bibitem{ham3}  J.A. Wheeler,  \textrm{Rev. Mod. Phys.} \textbf{29}, 463 (1957); \textrm{Ann. Phys. NY} \textbf{2}, 604 (1957); \textrm{in} \textit{Relativity, Groups and Topology}, ed. C. DeWitt and B. DeWitt, Gordon and Breach (1964), p.317; \textit{Einsteins Vision. Wie steht es beute mit Einsteins Vision, alles als Geometrie aufzufassen?}, Springer-Verlag Berlin Heidelberg New York, New York (1968); \textit{Geometrodynamics}, Academic Press, New York (1962).
\bibitem{ham4}  P.A.M. Dirac, \textrm{Proc. Roy. Soc. Lond. A} \textbf{246}, 326 (1958); \textit{ibid.} \textbf{246}, 333 (1958); \textrm{Phys. Rev} \textbf{114}, 924 (1959).
\bibitem{ham5}  P.W. Higgs, \textrm{Phys. Rev. Lett} \textbf{1}, 373 (1958); \textit{ibid.} \textbf{3}, 66 (1959).
\bibitem{ham6}  R. Arnowitt, S. Deser, and Ch.W. Misner, in \textit{Gravitation: an introduction to current research}, ed. L. Witten, John Wiley and Sons (1962), p.227, arXiv:gr-qc/0405109; \textrm{Phys. Rev.} \textbf{116}, 1322 (1959); \textrm{Phys. Rev} \textbf{117}, 1595 (1960); \textrm{J. Math. Phys} \textbf{1}, 434 (1960).
\bibitem{ham7}  A. Peres, \textrm{Nuovo Cim.} \textbf{26}, 53 (1962).
\bibitem{ham8}  R.F. Beierlein, D.H. Sharp, and J.A. Wheeler, \textrm{Phys. Rev.} \textbf{126}, 1864 (1962).
\bibitem{ham9}  A.B. Komar, \textrm{Phys. Rev.} \textbf{153}, 1385 (1967); \textit{ibid.} \textbf{164}, 1595 (1967).
\bibitem{ham10} B.S. DeWitt, \textrm{Gen. Rel. Grav.} \textbf{1}, 181 (1970).
\bibitem{ham11} V. Moncrief and C. Teitelboim, \textrm{Phys. Rev D} \textbf{6}, 966 (1972).
\bibitem{ham12} A.E. Fischer and J.E. Marsden, \textrm{J. Math. Phys.} \textbf{13}, 546 (1972).
\bibitem{ham13} C. Teitelboim, \textrm{Ann. Phys. NY} \textbf{80}, 542 (1973).
\bibitem{ham14} A. Ashtekar and R. Geroch, \textrm{Rep. Progr. Phys.} \textbf{37}, 1211 (1974).
\bibitem{ham15} T. Regge and C. Teitelboim, \textrm{Phys. Lett B} \textbf{53}, 101 (1974); \textrm{Ann. Phys. NY} \textbf{88}, 286, (1974).
\bibitem{ham16} K. Kucha$\mathrm{\check{r}}$, \textrm{J. Math. Phys.} \textbf{13}, 768 (1972); \textit{ibid.} \textbf{15}, No.6, 708 (1974).
\bibitem{ham17} C.J. Isham, \textrm{in} \textit{Quantum Gravity}, Oxford Symposium, eds. C.J. Isham, R. Penrose, and D.W. Sciama, Clarendon Press, Oxford (1975).
\bibitem{ham18} S.A. Hojman, K. Kucha$\mathrm{\check{r}}$, and C. Teitelboim, \textrm{Ann. Phys. NY} \textbf{96}, 88 (1976).
\bibitem{ham19} G.W. Gibbons and S.W. Hawking, \textrm{Phys. Rev. D} \textbf{15}, 2752, (1977).
\bibitem{ham20} D. Christodoulou, M. Francaviglia, and W.M. Tulczyjew, \textrm{Gen. Rel. Grav.} \textbf{10}, 567 (1979).
\bibitem{ham20a} M. Francaviglia, \textrm{Riv. Nuovo Cim.} \textbf{1}, 1 (1979).
\bibitem{ham20b} J.A. Isenberg, \textrm{in} \textit{Geometrical and topological methods in gauge theories}, \textrm{Lect. Notes Phys.} \textbf{129}, eds. J.P. Harnad and S. Shnider, Springer--Verlag, New York (1980).
\bibitem{ham21} J.A. Isenberg and J.M. Nester, \textrm{in} \textit{General Relativity and Gravitation. One Hundred Years After the Birth of Albert Einstein.}, ed. A. Held, Plenum Press, NewYork and London (1980), p.23
\bibitem{ham22} K. Kucha$\mathrm{\check{r}}$, \textrm{Phys. Rev. D} \textbf{39}, 2263 (1989).
\bibitem{ham23} Z. Bern, S.K. Blau, and E. Mottola, \textrm{Phys. Rev. D} \textbf{33}, 1212 (1991).
\bibitem{ham24} P.O. Mazur, \textrm{Phys. Lett B} \textbf{262}, 405 (1991).
\bibitem{ham25} C. Kiefer and T.P. Singh, \textrm{Phys. Rev. D} \textbf{44}, 1067 (1991).
\bibitem{ham26} C. Kiefer, \textrm{in} \textit{Canonical Gravity: From Classical to Quantum}, ed. J. Ehlers and H. Friedrich, Springer, Berlin (1994), arXiv:gr-qc/9312015
\bibitem{ham27} D. Giulini and C. Kiefer, \textrm{Class. Quantum Grav.} \textbf{12}, 403 (1995).
\bibitem{ham28} N. Pinto-Neto and E.S. Santini, \textrm{Phys. Rev. D} \textbf{59}, 123517 (1999).
\bibitem{ham29} N. Pinto-Neto and E.S. Santini, \textrm{Gen. Rel. Grav.} \textbf{34}, 505 (2002).
\bibitem{ham30} M.J.W. Hall, K. Kumar, and M. Reginatto, \textrm{J. Phys A: Math. Gen.} \textbf{36}, 9779 (2003).
\bibitem{ham31} N. Pinto-Neto, \textrm{Found. Phys.} \textbf{35}, 577 (2005).
\bibitem{ham32} M.J.W. Hall, \textrm{Gen. Rel. Grav.} \textbf{37}, 1505 (2005).
\bibitem{ham33} R. Carroll, \textrm{Theor. Math. Phys.} \textbf{152}, 904 (2007).
\bibitem{dir}   P.A.M. Dirac, \textrm{Can. J. Math.} \textbf{2}, 129 (1950); \textrm{Phys. Rev.} \textbf{114}, 924 (1959).
\bibitem{fad}   L.D. Faddeev, \textrm{Sov. Phys. Usp.} \textbf{25}, 132 (1982); \textrm{Usp. Fiz. Nauk} \textbf{136}, 435 (1982).
\bibitem{whe}   J.A. Wheeler, in \textit{Battelle Rencontres: 1967  Lectures in Mathematics and Physics}, eds. C.M. DeWitt and J.A. Wheeler (1968), p. 242
\bibitem{qgr1}  P. Gusin, \textrm{Phys. Rev. D} \textbf{77}, 066017 (2008).
\bibitem{qgr2}  T.P. Shestakova, \textrm{in} \textit{Proceedings of Russian summer school-seminar on Gravitation and Cosmology "GRACOS-2007"}, Kazan (2007), p.179, arXiv:0801.4854v1 [gr-qc]
\bibitem{a-v}   I.Ya. Aref'eva and I. Volovich, arXiv:0710.2696  [hep-ph]
\bibitem{qgr3}  Ch. Soo, \textrm{Class. Quantum Grav.} \textbf{24}, 1547 (2007), arXiv:gr-qc/0703074
\bibitem{qgr4}  D. Rickles, \textrm{in} \textit{The structural foundations of quantum gravity}, ed. D. Rickles, S. French, and J. Saatsi, Clarendon Press (2006), p.152
\bibitem{qgr5}  A.B. Henriques, \textrm{Gen. Rel. Grav.} \textbf{38}, 1645 (2006), arXiv:gr-qc/0601134
\bibitem{qgr6}  T. Kubota, T. Ueno, and N. Yokoi, \textrm{Phys. Lett. B} \textbf{579}, 200 (2004), arXiv:hep-th/0310109
\bibitem{qgr7}  K. Meissner, \textrm{Class. Quantum Grav.} \textbf{21}, 5245 (2004), arXiv:gr-qc/0407052
\bibitem{qgr8}  A. Ashtekar, M. Bojowald, and J. Lewandowski, \textrm{Adv. Theor. Math. Phys.} \textbf{7}, 233 (2003), arXiv:gr-qc/0304074
\bibitem{qgr9}  E. Anderson, J. Barbour, B. Foster, and N. 'O Murchadha, \textrm{Class. Quantum Grav.} \textbf{20}, 1571 (2003), arXiv:gr-qc/0211022
\bibitem{qgr10} G.F.R. Ellis, \textrm{in} \textit{Modern Cosmology}, ed. S. Bonometto, V. Gorini, U. Moschella (2002), ch.3
\bibitem{qgr11} C. Kiefer, \textrm{in} \textit{Towards Quantum Gravity: Proceedings of the XXXV International Winter School on Theoretical Physics, Held in Polanica, Poland, 2-11 February 1999}, \textrm{Lect. Notes Phys.} \textbf{541}, ed. J. Kowalski-Glikman, Springer (2000), p.158
\bibitem{qgr12} J.W. Norbury, \textrm{Eur. J. Phys.} \textbf{19}, 143 (1998), arXiv:physics/9806004
\bibitem{qgr13} A.O. Barvinsky and C. Kiefer, \textrm{Nucl. Phys. B} \textbf{526}, 509 (1998), arXiv:gr-qc/9711037
\bibitem{qgr14} T. Horiguchi, \textrm{Nuovo Cim. B} \textbf{112}, 1107 (1997); \textit{ibid.} \textbf{112}, 1227 (1997).
\bibitem{qgr15} N.P. Landsman, \textrm{Class. Quantum Grav.} \textbf{12}, L119 (1995), arXiv:gr-qc/9510033
\bibitem{qgr16} S. Carlip, \textrm{Class. Quantum Grav.} \textbf{11}, 31 (1994), arXiv:gr-qc/9309002
\bibitem{qgr17} D. Giulini, C. Kiefer, \textrm{Phys. Lett. A} \textbf{193}, 21 (1994).
\bibitem{qgr18} P. Mansfield, \textrm{Nucl. Phys. B} \textbf{418}, 113 (1994).
\bibitem{qgr19} M.D. Pollock, \textrm{Int. J. Mod. Phys. A} \textbf{7}, 4149 (1992).
\bibitem{qgr20} G. Hayward and K. Wong, \textrm{Phys. Rev. D} \textbf{46}, 620 (1992).
\bibitem{qgr21} A. Vilenkin, \textrm{Phys. Rev. D} \textbf{39}, 1116 (1989).
\bibitem{qgr22} S. Weinberg, \textrm{Rev. Mod. Phys.} \textbf{61}, 1 (1989).
\bibitem{qgr23} M. McGuigan, \textrm{Phys. Rev. D} \textbf{38}, 3031 (1988).
\bibitem{qgr24} S.W. Hawking, \textrm{Nucl. Phys. B} \textbf{239}, 257 (1984).
\bibitem{qgr25} J.B. Hartle and S.W. Hawking, \textrm{Phys. Rev. D} \textbf{28}, 2960 (1983).
\bibitem{qgr26} B.S. DeWitt, \textrm{Phys. Rep.} \textbf{19}, 295 (1975).
\bibitem{qgr27} P.G. Gilkey, \textrm{J. Diff. Geom.} \textbf{10}, 601 (1975); \textrm{Proc. Symp. Pure. Math.} \textbf{27}, 265 (1975).
\bibitem{qgr28} H.P. McKean and I.M. Singer, \textrm{J. Diff. Geom.} \textbf{5}, 233 (1971).
\bibitem{qgr29} B.S. DeWitt, \textit{Dynamical Theory of Groups and Fields}, Gordon and Breach (1965).
\bibitem{sup}   A.E. Fischer, \textrm{in} \textit{Relativity}, eds. M. Carmeli, S.I. Fickler, and L. Witten, Plenum Press, New York (1970), p. 303; \textrm{Gen. Rel. Grav} \textbf{15}, 1191 (1983); \textrm{J. Math. Phys} \textbf{27}, 718 (1986).
\bibitem{min1}  S.W. Hawking, \textrm{in} \textit{Pontificiae Academiae Scientiarum Scripta Varia} \textbf{48}, 563 (1982).
\textrm{in} \textit{Relativity, Groups and Topology II}, Les Houches 1983, Session XL, eds. B.S. DeWitt and R. Stora, North Holland, Amsterdam (1984), p. 333;
\textrm{in} \textit{300 Years of Gravitation}, eds. S.W. Hawking and W. Israel, Cambridge University Press, Cambridge (1987), p. 631;
\textrm{Phys. Rev. D} \textbf{32}, 2489 (1985).
\bibitem{min2}  A. Linde, \textrm{Rep. Prog. Phys.} \textbf{47}, 925 (1984).
\bibitem{min3}  R. Brandenburger, \textrm{Rev. Mod. Phys.} \textbf{57}, 1 (1985).
\bibitem{min4}  J.J. Halliwell and S.W. Hawking, \textrm{Phys. Rev. D} \textbf{31}, 1777 (1985).
\bibitem{min5}  S.W. Hawking and J.C. Luttrell, \textrm{Phys. Lett. B} \textbf{143}, 83 (1984); \textrm{Nucl. Phys. B} \textbf{247}, 250 (1984).
\bibitem{min6}  D. Page, \textrm{Phys. Rev. D} \textbf{32}, 2496 (1985).
\bibitem{min7}  P. Amsterdamski, \textrm{Phys. Rev. D} \textbf{31}, 3073 (1985).
\bibitem{ish}   C.J. Isham, \textrm{in} \textit{Quantum Theory of Gravity. Essays in honor of the 60th birthday of Bryce S. De Witt}, eds. S.M. Christensen and Adam Hilger, Bristol (1984), p. 299
\bibitem{mtw}   C.W. Misner, K.S. Thorne, and J.A. Wheeler, \textit{Gravitation}, W.H. Freeman and Company, San Francisco (1973).
\bibitem{neu}   J. von Neumann, \textrm{Math. Ann.} \textbf{104}, 570 (1931).
\bibitem{a-w}   H. Araki and E.J. Woods, \textrm{J. Math. Phys.} \textbf{4}, 637 (1963).
\bibitem{bos}   J.-P. Blaizot and G. Ripka, \textit{Quantum theory of f\/inite systems}, Massachusetts Institute of Technology Press, Cambridge (1986).
\bibitem{ber}   F.A. Berezin, \textit{The Method of Second Quantization}, 2nd ed., Nauka, Moscow (1987).
\bibitem{bs}    N.N. Bogoliubov and D.V. Shirkov, \textit{Introduction to the theory of quantized fields}, 3rd ed., John Wiley and Sons, (1980)
\bibitem{bog}   N.N. Bogoliubov, A.A. Logunov, A.I. Oksak, and I.T. Todorov, \textit{General Principles of Quantum Field Theory}, Nauka, Moscow (1991).
\bibitem{veltman} M.J.G. Veltman, \textrm{in} \textit{Methods in Field Theory. Proc. Les Houches, Session XXVIII},  eds. R. Balian and J. Zinn-Justin, North Holland (1976), p. 265
\bibitem{th1}   T. Horiguchi, \textrm{Mod. Phys. Lett. A} \textbf{8}, 777 (1993); \textrm{Phys. Rev. D} \textbf{48}, 5764 (1993).
\bibitem{th2}   M.J. Duncan, \textrm{Nucl. Phys. B} \textbf{361}, 695 (1991).
\bibitem{th3}   W. Fishler, I. Klebanov, J. Polchinski, and L. Susskind, \textrm{Nucl. Phys. B} \textbf{327}, 157 (1989).
\bibitem{th4}   S.B. Giddingsa and A. Strominger, \textrm{Nucl. Phys. B} \textbf{321}, 481 (1989).
\bibitem{th5}   A. Hosoya and M. Morikawa, \textrm{Phys. Rev. D} \textbf{39}, 1123 (1989).
\bibitem{th6}   M. McGuigan, \textrm{Phys. Rev. D} \textbf{38}, 3031 (1988); \textit{ibid.} \textbf{39}, 2229 (1989).
\bibitem{th7}   V.A. Rubakov, \textrm{Phys. Lett. B} \textbf{214}, 503 (1988).
\bibitem{glinka} L.A. Glinka, \textit{{\AE}thereal Multiverse: A New Unifying Theoretical Approach to Cosmology, Particle Physics, and Quantum Gravity}, Cambridge International Science Publishing, Great Abington, UK (2012).
\end{thebibliography}
\end{document}